\documentclass[%
 reprint,
 amsmath,amssymb,
 aps,
]{revtex4-1}

\usepackage{graphicx}% Include figure files
\usepackage{dcolumn}% Align table columns on decimal point
\usepackage{bm}% bold math
\begin{document}

\preprint{APS/123-QED}

\title{Long valley lifetime of free carriers in monolayer WSe$_2$}% Force line breaks with \\

\author{Tengfei Yan}
\author{Siyuan Yang}%
\author{Dian Li}
\author{Xiaodong Cui}
\email{e-mail: xdcui@hku.hk}
\affiliation{%
 Department of Physics, the University of Hong Kong\\
}%

\date{\today}% It is always \today, today,

\begin{abstract}
	Monolayer transition metal dichalcogenids (TMDs) feature valley degree of freedom, giant spin-orbit coupling and spin-valley locking. These exotic natures stimulate efforts of exploring the potential applications in conceptual spintronics, valleytronics and quantum computing. Among all the exotic directions, a long lifetime of spin and/or valley polarization is critical. The present valley dynamics studies concentrate on the band edge excitons which predominates the optical response due to the enhanced Coulomb interaction in two dimensions. The valley lifetime of free carriers remains in ambiguity. In this work, we use time-resolved Kerr rotation spectroscopy to probe the valley dynamics of excitons and free carriers in monolayer tungsten diselinide. The valley lifetime of free carriers is found around 2 ns at 70 K, about 3 orders of magnitude longer than the excitons of about 2 ps. The extended valley lifetime of free carriers evidences that exchange interaction dominates the valley relaxation in optical excitation. The pump-probe spectroscopy also reveals the exciton binding energy of 0.60 eV in monolayer WSe$ _2 $.
\begin{description}
\item[PACS numbers]
78.66.Li, 72.25.Rb, 71.35.Cc.
\end{description}
\end{abstract}

\pacs{Valid PACS appear here}% PACS, the Physics and Astronomy
\maketitle

%\tableofcontents

%\section{}

In solid state physics, valley refers to the local energy extreme, either conduction band local minimum or valence band local maximum, in crystal electronic band structures. The occupation of carriers at inequivalent valleys, carrying different momentum phase, represents different quantum states. This leads to the conceptual valleytronics, which utilizes the valley degree of freedom as a quantum information carrier, in a similar way as the spintronics where spin degree of freedom is utilized.\cite{rycerz2007valley,xiao2007valley,shkolnikov2002valley,xiao2010berry} Monolayer transition metal dichalcogenides (TMDs), the emerging 2D semiconductor, features degenerate but inequivalent valleys K and K' (or -K) located at band edges of both conducting band and valence band, which are separated by a big momentum space. Owing to the spatial inversion symmetry breaking in monolayer TMDs, the Berry curvature, a function describing properties of valence electron orbits in crystal lattices, shows opposite signs at K and K' valleys. This could work as a knob to selectively manipulate the K or K' valley.\cite{xiao2012coupled,zeng2012valley,cao2012valleyselective,mak2012control,sallen2012robust} Besides, the K and K' valleys are constructed by metal's d orbits which experience strong spin-orbit coupling (SOC). SOC splits the band, particularly valence band, into two sub-bands. As a result of time reversal symmetry, the spin splitting shows opposite signs between K and K' valleys at equal energies.\cite{xiao2012coupled,zeng2013optical,yuan2013zeeman,gong2013magnetoelectric} Namely, if the band edge at K valley is  spin-up state, the band edge at K' valley must be spin-down as illustrated in FIG. 1(a). This leads to a definite relationship between valley and spin indices, the so called spin-valley locking.\cite{srivastava2015valley,macneill2015breaking,aivazian2015magnetic} The unique spin-valley locking interplays the valley and spin degrees of freedom and suppresses the valley/spin relaxation: In the relaxation process of a hot carrier around K valley, the conservation of momentum (valley) and spin must be both satisfied. It theoretically supports a long valley/spin lifetime in monolayer TMDs.

\begin{figure*}[!htb]
	\centering
	\includegraphics[width=0.8\textwidth]{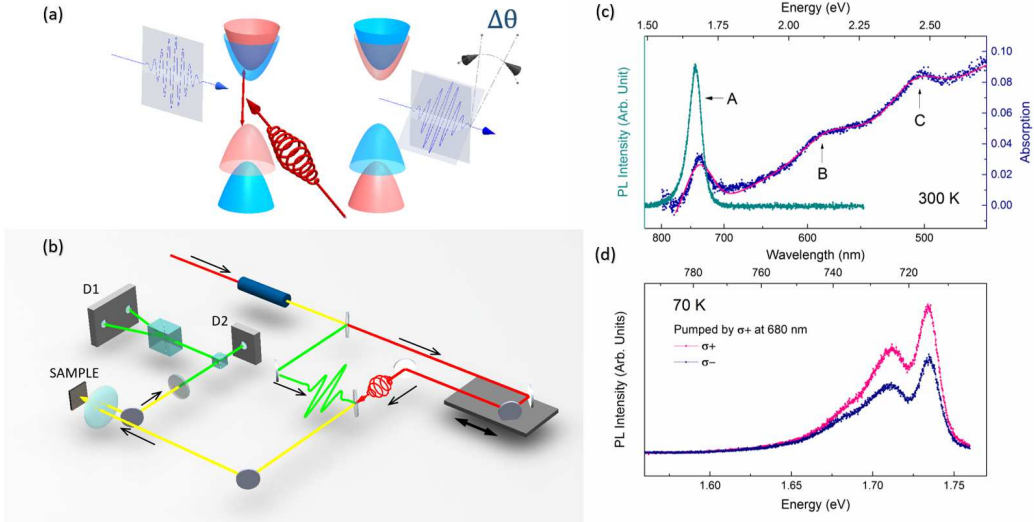}
	\caption{(a) A schematic diagram of the monolayer WSe$ _2 $ band structure in K (left) and K' (right) valleys, their spin states (shown in red and blue colors, respectively) and the light-matter interaction in the time-resolved Kerr rotation experiment. (b) Schematic of the pump-probe Kerr rotation spectroscopy. D1 is a balanced photo-detector and D2 a photodiode, working together to get the Kerr rotation as a optical bridge. D2 is a photodiode monitoring the reflectivity change. (c) PL spectrum of the monolayer WSe$_2$ on Si/SiO$_2$ substrate (green dots), and absorption spectrum on mica (the blue dots) with a Gaussian function fit (red line). The data are measured at room temperature. (b) Helicity-resolved PL spectra of monolayer WSe$_2$ pumped by a 1.824 eV left-handed circularly polarized laser ($\sigma+$) at 70 K.}
\end{figure*}

\vspace{18 pt}
On the experimental side, however, the valley/spin lifetime of free carriers  remains some ambiguity. A spin resolved photocurrent measurements estimated the valley/spin lifetime in the range of $ 10^0 \sim 10^2 $ nanoseconds in monolayer WS$ _2 $, while optical pump-probe spectroscopy and time-resolved photoluminescence (PL) experiments gave a very short valley lifetime of several picoseconds\cite{wang2013valleycarrierdynamics,mai2014manybodyeffects,kumar2014valley,plechinger2014time,lagarde2014carrier,wang2014valleydynamics,zhu2014excitonvalley,plechinger2016trion} with a few exceptions where long valley lifetime of bound excitons were reported.\cite{yang2015long,yang2015spin,hsu2015optically} The huge discrepancy lies in that the excitonic effect is prevalent in optical responses of monolayer TMDs.\cite{qiu2013opticalspectrum,he2014tightly,zhu2015excitonbinding,ye2014probingexcitonic,chernikov2014excitonbinding,ugeda2014giant,zhang2014direct} The giant exciton binding energy implies a short effective radius of excitons, a close separation between electrons and holes, enhancing the  exchange interactions. Spin exchange interactions are believed to be the major valley/spin depolarization channel in monolayer TMDs that causes the short valley/spin lifetime of excitons.\cite{jones2013opticalcoherence,yu2014valleydepolarization,zhu2014excitonvalley,glazov2014excitondecoherence}  To date, the direct measurement of free carriers is lacking. In this report, we use time-resolved Kerr rotation spectroscopy to unambiguously identify the valley/spin lifetime of free carriers in monolayer WSe$_2 $.  It is 3 orders of magnitude longer than that of excitons. The results show monolayer TMD is a promising platform for conceptual valleytronics and non-magnetic spintronics.

The WSe$_2$ flakes are mechanically exfoliated from single crystal WSe$_2$ onto Si/SiO$_2$ substrate for PL and pump-probe measurements, and mica for transmittance measurement. The two-color pump-probe measurement set-up is shown in FIG. 1(b).  More details of the helicity-resolved PL and the pump-probe measurement set-ups have been illustrated elsewhere.\cite{zeng2012valley,yan2015valleydepolarization,yan2015excitonvalleydynamics} Monolayer WSe$ _2 $ flakes are identified by optical mircoscope and PL.

%\section{Results}

The optical interband transitions are characterized by PL and absorption spectroscopy at room temperature as shown in FIG. 1(c). There are three prominent absorption peaks, in consistent with previous reports,\cite{zhao2012evolution} which are labeled as A, B and C. Peak A and B correspond to the two exciton states of spin-split interband transitions at K and K' valley. Peak C corresponds to the several exciton states of interband transition near $\Lambda$ and $ \Gamma $ point in the Brillouin zone as a result of the band nesting effect.\cite{carvalho2013bandnesting,kozawa2014photocarrier} The energy shift of A exciton between the absorption and PL spectra is presumably due to the different exciton binding energy modified by the different dielectric permittivity of the mica and SiO$_2$ substrates.\cite{ugeda2014giant}

FIG. 1(d) shows the helicity resolved PL spectra of a WSe$_2$ flake excited by a $\sigma+$ laser at 70 K.  The emission peak at 1.734 eV and the lower energy peak at 1.712 eV are identified as the A excitons and the charged excitons (trions). The collected PL signal shows prominent circular polarization about 24\% for both excitons and trions, indicating a clear valley polarization in K and K' valleys. The lower energy tail consisting of defect bound excitons shows negligible valley polarization.

\begin{figure}[!htb]
	\centering
	\includegraphics[width=0.5\textwidth]{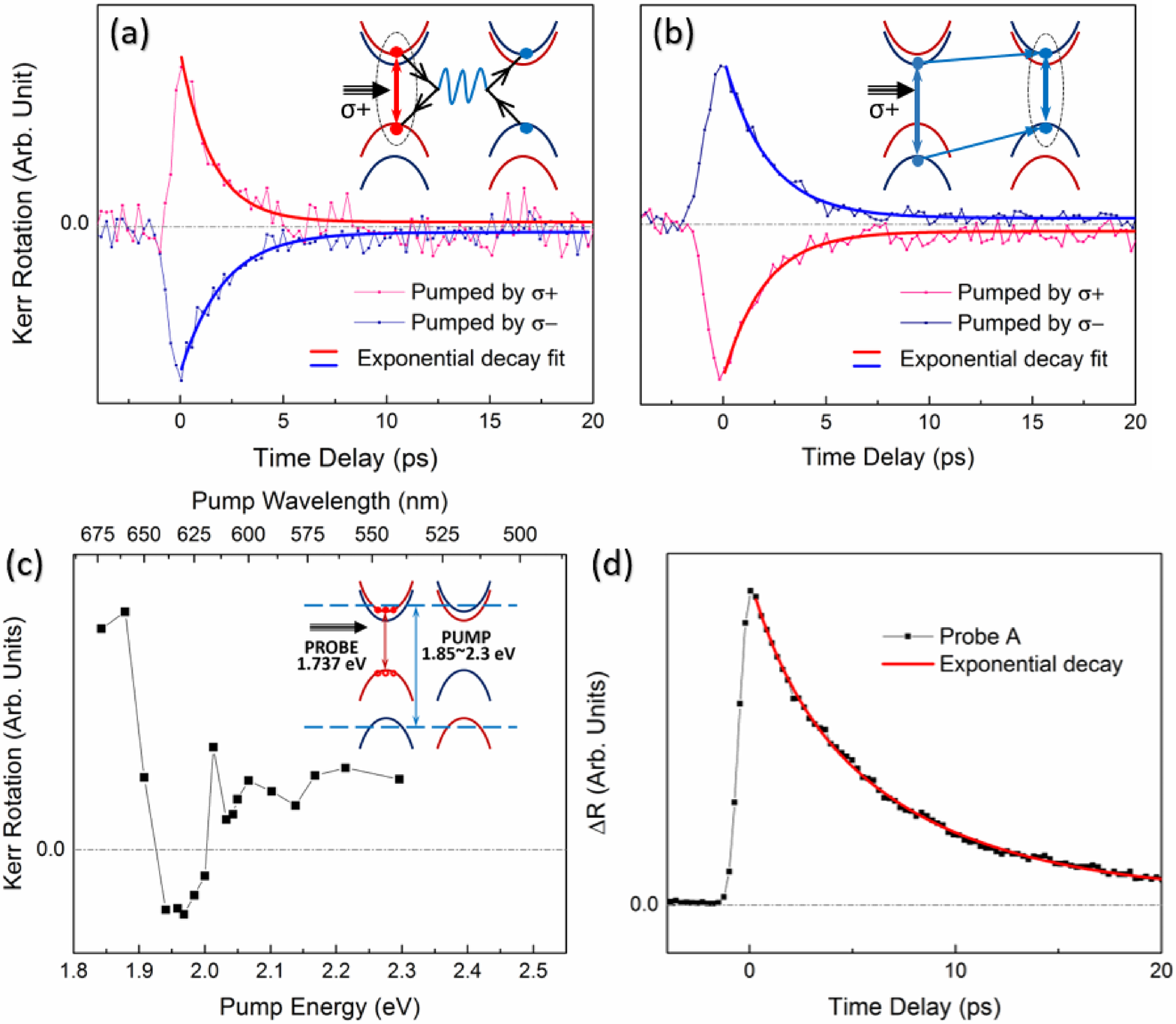}
	\caption{(a,b) Time-resolved Kerr rotation of monolayer WSe$_2$ measured at 70 K. The red and blue dotted lines indicate the Kerr rotation traces when pumped by left-handed (red) and right-handed (blue) circularly polarized pulses, respectively. The solid lines follow an exponential decay fit. (a) The sample is pumped by the pulse centered at 1.834 eV and probed at 1.737 eV (pumping A, probing A). The insets illustrate band structure and transitions between different bands. The valley depolarization is dominated by electron-hole spin exchange. (b) The sample is pumped at 2.023 eV and probed at 1.737 eV (pumping B, probing A). The sign change in Kerr rotation signal implies that the B excitons first relax to A excitons via intervalley scatterings, shown in the inset. The similar decay slope to (a) indicates the same depolarization channel. (c) The Kerr rotation angel at zero time delay as a function of the pumping energy. (d) The time-resolved reflectance spectrum probed at 1.737 eV and pumped at 1.834 eV, with an exponential decay function fit in red solid line.}
\end{figure}

We measured the valley relaxation of A excitons by pumping the sample at 1.834 eV and probing at 1.737 eV (pumping A, probing A). The probing beam intensity is tuned to be a tenth of the pumping beam to minimize its influence. The pumping beam injected exciton density is estimated to be at the magnitude of $ 10^{12} \ \rm{cm}^{-2} $, given the absorption ratio shown in FIG. 1(c) and the laser spot radius of 1 $ \rm{\mu m} $. Both data with left-handed and right-handed circularly polarized pumping beams are shown in FIG. 2(a). The valley polarization relaxation are well described with single exponential decay functions. The valley lifetime is extracted to be 1.8$\pm$0.3 ps, in consistent with the previous studies.\cite{zhu2014excitonvalley,yan2015excitonvalleydynamics} Such a quick depolarization process is attributed to exciton intervalley (K-K') scattering through the strong electron-hole exchange interaction, which is diagrammed in the inset of FIG. 2(a).\cite{jones2013opticalcoherence,yu2014valleydepolarization} The time-resolved reflection spectrum with the same pumping and probing energy is shown in FIG. 2(d). The exponential decay function fit shows $ \Delta R $ decay time constant is 7$ \pm $0.2 ps, which may related to the phase space redistribution via phonon-exciton and exciton-exciton scatterings, longer than the valley relaxation time.\cite{pollmann2015resonant} The exciton formation time is deduced to be less than 0.5 ps, limited by the time resolution of the experiment set-up used here. 

\begin{figure}[!htb]
	\centering
	\includegraphics[width=0.5\textwidth]{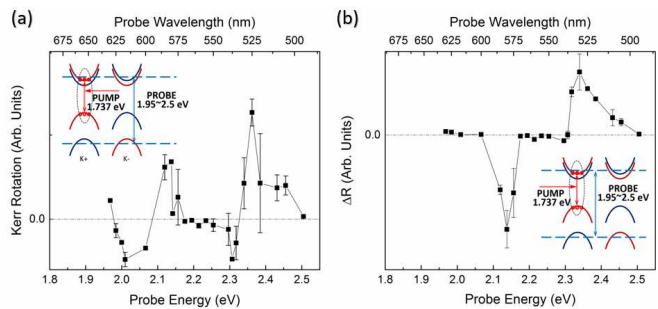}
	\caption{(a)(b) The transient differential reflection intensity and the Kerr rotation angel at zero time delay as a function of the probing energy. Both experiments are conducted with the pumping laser energy set at 1.737 eV and the probing laser energy tuned from 1.95 to 2.5 eV. }
\end{figure}

While we tune the pumping energy to 2.023 eV, near resonant with B excitons, the Kerr rotation at 1.737 eV (pumping B, probing A) changes its sign as shown in FIG. 2(b). It implies that the spin conserved intervalley scattering prevails over the spin-flip intravalley scattering in the hot carrier relaxation process, as illustrated in the inset of FIG. 2(b). For B excitons generated by left-handed circularly polarized pump in K valley, the holes could be scattered to the valence band edge and electron to the higher spin-split sub-band of conduction band at the K' valley via Coulomb interactions without spin flip, forming A excitons at the K' valley. Thus a sign change in Kerr rotation is observed. The rise time of the Kerr rotation signal here is approximately 0.8 ps longer than that in the case of pumping A probing A, possibly resulting from the time of hot exciton intravalley relaxation, intervalley scattering and formation of A excitons. The relaxation time is deduced to be 2$ \pm $0.2 ps, same as that of pumping A probing A situation within the error bar. It is consistent with the proposed relaxation mechanism, for the depolarization process of the A excitons in both cases shares the same relaxation channel, the electron-hole exchange interactions. 

The Kerr rotation angle measured at zero time delay is plotted as a function of the pumping energy, which is shown in FIG. 2(c). There is a clear region centered at 1.97 eV that the Kerr rotation exhibits a negative signal. The region is assigned to the B exciton. When assuming that the energy temperature dependence and the binding energies are similar for the A, B excitons, FIG. 1(c) suggests the energy of B exciton around 2.184 eV based on the steady-state optical measurements. The large energy shift of about 0.21 eV is attributed to the bandgap renormalization, i.e., the energy lowering correlation of free carriers, caused by the dense exciton density up to  $ 10^{12} \ \rm{cm^{-2}} $ injected by the pumping pulses in the transient measurement. The energy shift is much larger compared to the quasi-2D systems like GaAs quantum wells.\cite{kleinman1985band,trankle1987dimensionality,wang2013valleycarrierdynamics}

\begin{figure*}[!htp]
	\centering
	\includegraphics[width=1\textwidth]{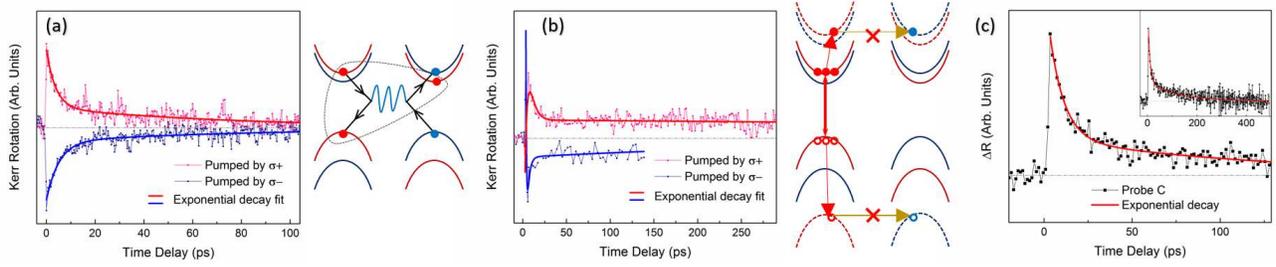}
	\caption{(a) Time-resolved Kerr rotation of monolayer WSe$_2$ pumped at 1.8 eV and probed at 1.71 eV (trion) at 70 K. The illustration of negative charged exciton valley relaxation is shown at the right side. (b) Kerr rotation pumped at 1.737 eV and probed at 2.318 eV at 70 K, and the sketch of band structure with the dashed lines indicate the quasi-particle bands without the modification of excitonic effects. (c) The time-resolved reflectance pumped at 1.737 eV and probed at 2.318 eV. The inset shows same trace in a larger time scale.}
\end{figure*}

To further study the valley dynamics in monolayer WSe$ _2 $, we set the pumping energy to be 1.737 eV, which is resonant with the A exciton, and tune the probing energy in the range between 1.968 and 2.505 eV. The transient differential reflection at zero time delay are plotted as a function of the probing energy shown in FIG. 3(b). Negative differential reflection peaking at 2.15 eV is observed, close to 2.184 eV of B excitons extracted from the steady-state optical measurements. The transition is confirmed at the same energy in FIG. 3(a), in which the probing energy dependent Kerr rotation spectrum at zero time delay is shown. The redshift of 0.034 eV of B exciton is remarkable different from that of 0.21 eV in FIG. 2(c), owing to bandgap renormalization concentrating on different states. In the set-up shown in FIG. 3, the bandgap renormalization affects most on the A exciton states (pumping resonantly A excitons) and only few excitons are scattered to B states; while in the set-up in FIG. 2(c), hot carriers concentrate on states around B excitons (pumping resonantly B excitons). 

The differential reflection spectrum (FIG. 3(b)) also reveals another transition when the probing energy is around 2.32 eV. A typical $ \Delta R $ curve is shown in FIG. 4(c). The positive differential reflection usually results from the band filling effect,  implying that the transition is likely associated with the continuum state of A exciton, i.e., the free carriers from the band edge transition. A portion of the pumped A excitons could be excited to higher excited states and eventually be ionized to free carriers due to the exciton-exciton annihilation, phonon scattering or localized electric field formed by defects. The ionized electrons and holes occupy the quasi-particle electronic band edges. Note that the population of C excitons upon pumping around A exciton states is negligible owing to the large energy separation.\cite{kozawa2014photocarrier} This is evidenced by the time-resolved reflectance spectra. A typical trace is shown in FIG. 4(c), in which the relaxation process is fitted by an exponential decay function. The signal decays in 130$ \pm $13 ps after a rapid decay characterized by a time constant of 8$ \pm $1 ps. The fast component, similar and may be related to the relaxation of excitons shown in FIG. 2(d) via exciton formation. The slow component indicates that the reduced oscillator strength of ionized excitons results in weaker recombination and scatterings. This excludes possibility that the signal originates from the response of C excitons, which is expected to be extremely fast because of intraband relaxation.

The probing energy dependent Kerr rotation spectrum at zero time delay shown in FIG. 3(a) reads significant Kerr signal across that area. This also rules out the potential detection of high energy excitons (C excitons) around $ \Gamma $ points in their Brillouin zone as valley dependent optical selections is not valid around the Brillouin zone center. So it is concluded that the transition around 2.32 eV corresponds to the quasi-particle band edge. Compared to the PL result, the exciton binding energy is extracted to be 0.596 eV, close to the values reported by other methods.\cite{wang2015giant,hanbicki2015measurement}

FIG. 4 shows the time-resolved Kerr rotation spectra of trions and free carriers under pumping A excitons. The trion valley polarization (probing at 1.71 eV) relaxes significantly slower than that of A excitons, qualitatively consistent with the previous reports.\cite{yang2015long,yang2015spin,hsu2015optically} A two-section exponential fit gives time constants of $ 5 \pm 1 $ ps and $ 80 \pm 14 $ ps, respectively.  The valley polarization of free carriers (probing at 2.318 eV) experiences a rapid decay near zero time delay followed by an even slower decay, with time constants of $ 5\pm 2 $ ps and $ 2.4\pm 1$ ns, respectively. The large error results from the relatively short delay line limited by the experiment conditions. The 3 orders of magnitude increase of the valley lifetime at free carrier states originates from the suppressed valley relaxation channel via electron-hole exchange interactions. The increased spatial separation between free electrons and holes dramatically weakens the spin exchange interactions which dominates the valley depolarization of excitons. Considering the electron-hole symmetry and the large spin-splitting in valence band, we attribute the Kerr rotation signal to the valley polarization of the holes. 

Note that both the rapid relaxation channel of trions and the continuum states share the similar fast decay process after initial pumping, with the similar time constant with A excitons. Thus we infer that the valley polarization of high-density trions and free carriers decays rapidly via band edge exciton states. It is suggested that the trion's valley depolarization likely goes through similar relaxation channel as excitons, electron-hole exchange interactions as illustrated in FIG. 4(a). The prolonged valley lifetime of trions likely results from the weaker Coulomb interaction compared to that in excitons, while this relaxation channel is invalid for free carriers.

\vspace{18 pt}

%\section{conclusion}

In conclusion, monolayer WSe$ _2 $ is examined by the time-resolved Kerr rotation technique. We have experimentally revealed the exciton binding energy to be 0.60 eV. The valley relaxation time constants of excitons, trions and free carriers are derived to be approximately 2 ps, 80 ps and 2 ns at 70 K, respectively. Our observations of the valley relaxation in monolayer WSe$ _2 $ provides a new insight for the valley dynamics in monolayer TMDs and valleytronics development. 

%\section*{Author Contributions}

%XDC conceived the experiments; TFY, SYY and DL performed the experiments; TFY and XDC analyzed the data and wrote the manuscript.

\providecommand{\noopsort}[1]{}\providecommand{\singleletter}[1]{#1}%

\clearpage
\nocite{*}

\bibliography{apssamp}% Produces the bibliography via BibTeX.

\end{document}